Title
**Influence of bone microstructure on ultrasound loss through skull-mimicking digital phantoms**

Authors
Samuel Clinard[1], Taylor Webb[2], Henrik Odéen[2], Dennis L. Parker[2,1], Douglas A. Christensen[1,3]

1. Department of Biomedical Engineering, University of Utah, Salt Lake City, Utah, USA

    Department of Radiology and Imaging Sciences, University of Utah, Salt Lake City, USA

2. Department of Electrical and Computer Engineering, University of Utah, Salt Lake City, USA

Sam.Clinard@utah.edu

729 Arapeen Dr.

3. Salt Lake City, Utah 84108

Abstract:
**Background:** Transcranial focused ultrasound treatments rely on precisely delivering ultrasound through the inhomogeneous human skull. Full-wave ultrasound simulations are a means to predict and correct the resulting ultrasound aberrations and attenuation. To do this, the acoustic properties of the skull, including phase velocity and attenuation, must be determined. A common approach relates computed tomography (CT) Hounsfield Units (HU) to these acoustic properties. In the trabecular regions of skulls, the CT HU values will depend on the fraction of bone and marrow within an image volume element, but they are typically insensitive to the microstructure of the bone and marrow.

**Purpose:** This study explores the influence of bone/marrow microstructures on determining the relationship of acoustic properties, particularly loss, to CT HUs. The typical clinical CT resolution (0.5 mm) cannot resolve fine trabecular bone microstructure, suggesting the relationship of attenuation to HU may be ill-determined.

**Methods:** The ultrasound insertion loss was found through various skull-mimicking digital phantoms consisting of two constituent materials (red marrow and cortical bone) from 0 to 75% porosity. The phantoms were assigned one of six pore diameters ranging from 0.2 mm to 1.0 mm. Ultrasound simulations were computed using k-Wave with a continuous 230 kHz or 650 kHz uniform pressure source. The insertion loss with and without absorption was defined as the mean pressure through the phantom with respect to the mean pressure in a water-only reference.

**Results:** The simulations at 230 kHz showed that the loss changed with porosity, but specific microstructure had little effect. However, in both non-absorbing and absorbing 650 kHz source simulations, the insertion loss depended on porosity and pore diameter. Larger pore diameter phantoms generally had higher losses than smaller pore diameter phantoms at the same porosity. In the non-absorbing phantoms, the maximum range in insertion loss was 2% to 52% over the range of pore diameters, which occurred at 20% porosity. Absorbing phantoms increased the loss by an average of 8.2%, with the greatest increase of 13% occurring for the smallest pore diameter at 2.5% porosity. Coherent multiple reflections from the phantom's planar interfaces influenced the loss within smaller pore diameter phantoms. The phase coherence of these reflections was disrupted by increased scattering within the larger pore diameter phantoms.

**Conclusion:** The results suggest that the relationship between attenuation and clinical HUs is ill-determined at 650 kHz, since the insertion loss depends on both porosity and pore diameter. The demonstrated uncertainty has important implications for developing CT-derived acoustic models of skull bone, as no single attenuation value can be related to HUs comprised of variable microstructures. Generally, our results show larger pore diameters (coarse microstructures) have higher loss than smaller pore diameters (fine microstructures) at the same porosity, which is consistent with scattering theory.

INTRODUCTION

Transcranial focused ultrasound is a noninvasive therapeutic modality used to treat various neurological disorders. The safety and efficacy of these procedures rely on precisely focusing the ultrasound beam through the inhomogeneous human skull. A fundamental challenge is accurately determining the acoustic properties of the skull in order to predict the beam intensity at the focus and to find the phase and amplitude compensation needed to adjust for the inhomogeneities.[1] A common approach relates the speed of sound and attenuation relationships to Hounsfield Units (HU) obtained by pretreatment computed tomographic (CT) imaging.[2–5] There are substantial differences between the several relationships that have been proposed, and disagreements exist on which to use clinically.[6,7]

The differences in the models may arise from several factors, including what type of experiments were used and what parametric functions were used to fit the HU values to attenuation.[2–4,6,8–12] Reviews and studies comparing various models have not found a consensus.[7,13] Several of the reported attenuation relationships are plotted together in Figure 6 of Leung et al. 2019, providing a helpful visualization of the large variability across studies.[6] Two approaches are commonly used to address this uncertainty. One method is simple derating based on ultrasound frequency, skull thickness, and a conservative (low) attenuation value for the skull.[14] Alternatively, full-wave simulations commonly use a single attenuation value for skull bone.[15] These simulations model the shape of the water-skull-brain interface and skull thickness; however, they do not directly model the internal skull structure. While safe, these conservative approaches potentially limit the delivery of higher ultrasound intensity, which may be more effective. As such, a better understanding of the ultrasound interactions with the internal skull structure may enable more precise simulations.

One difficulty is the limited resolution of clinical CT images in skull bone (typically 0.5 mm in plane), which cannot resolve the finer trabecular microstructure that contributes significantly to attenuation.[3,16–18] For this reason, the attenuation relationship to clinical CT HUs may be ill-determined. The HU of a given voxel is related to the average density in the voxel, which is strongly influenced by the porosity of the constituent materials and, to a lesser extent, the mineralization of the bone tissue.[19] The exact relationship of HUs to density depends on the bone type. While a rigorous study of HUs and density in skull bone has not been completed, many authors have assumed a linear relationship.[2,11,20] However, a given average density and corresponding HU may be associated with various microstructures due to the limited CT resolution.

Skull bone generally has three layers: two outer cortical layers and a middle trabecular layer. The porous trabecular bone consists primarily of cortical bone interspersed with blood vessels and red marrow pores and typically has porosities greater than 30%.[21,22] The cortical layers can range in porosity from 5% to 30%, while the trabecular layers can have porosities of 30% to 90%.[21] The distribution of these constituent materials, i.e., the bone microstructure, is expected to influence ultrasound attenuation.[16]

Previous work has explored ultrasound interactions with bone microstructure to diagnose osteoporosis. In that application, ultrasound attenuation, which varies with the bone mineral content and microstructure, has been shown to predict fracture risk.[23] Furthermore, the strong frequency dependence of absorption and scattering loss enables microstructure characterization with multi-frequency broadband sources.[24,25] While the attenuation's dependence on microstructure is helpful in interpreting diagnostic ultrasound, it causes challenges in transcranial therapeutic ultrasound. While it is known that there is an effect, the exact influence of skull microstructure in determining the skull's attenuation is still poorly understood.[3,18] Studies are limited that examine the impact of bone microstructure at frequencies relevant to therapeutic ultrasound (<1 MHz). tFurther, skull microstructure is significantly different from load-bearing bone.[26] In this study, we use methods developed for quantitative ultrasound to explore how the skull bone microstructure affects ultrasound loss at frequencies relevant to therapeutic ultrasound.

Quantitative ultrasound studies have used full-wave ultrasound simulation through simplified bone-mimicking digital phantoms to discern the microstructure's effect on attenuation.[24,27] The digital phantoms typically comprise cylindrical (3D) or circular (2D) marrow pores placed into a cortical background. Cylindrical pores best represent load-bearing bones, such as the calcaneus and femur, where the trabeculae are anisotropic plates and rods roughly oriented in the load-bearing direction.[28] Skull bone is non-load bearing; as such, it doesn't have cylindrical-like structures.[18,29] Our phantoms comprise spherical pores, which we believe better approximate the skull microstructure. This difference in microstructure likely affects the loss, as spherical and cylindrical scatterers are known to have different scattering properties.[30] Further, we allow our pores to overlap, creating more complex microstructures.

We then conducted full-wave ultrasound simulations through these digital phantoms to test whether features of the microstructure that clinical CT does not resolve can affect acoustic loss. We constructed digital phantoms with varying porosities (and associated HUs) but different pore sizes. We then simulated their insertion loss as a function of both pore size and porosity. Based on scattering theory, we hypothesized that larger pore sizes would lead to greater acoustic loss even when the porosity was kept constant.[31,32] Our results are consistent with this hypothesis, demonstrating that, at 650 kHz, acoustic loss in the skull depends on microstructure properties that CT HUs do not resolve.

METHODS

Bone Phantoms

Three-dimensional (12.8 mm x 12.8 mm x 5 mm) bone-mimicking digital phantoms were generated with varying porosities and pore diameters. Each phantom had an isotropic resolution of 0.05 mm and was constructed using only two constituent materials: marrow and cortical bone. The properties of each of these components, along with nominal values for the water that surrounds the phantom, are listed in Table I. The phantoms were constructed by randomly placing marrow pores of a fixed diameter into a cortical bone background. Pores were allowed to overlap to mimic the complex trabecular microstructure better. A total of six pore diameters were chosen (0.2 mm, 0.3 mm, 0.4 mm, 0.6 mm, 0.8 mm, 1.0 mm). Each pore diameter was used to create eight phantoms, with porosities ranging from 2.5% to 75%, for a total of 48 phantoms. Figure 1 shows cross sections of nine of these phantoms, illustrating three representative pore diameters at three porosities. Details of the phantom generation and randomization scheme are given in the supplemental materials. At each nominal porosity and pore size, five phantoms with different random pore positions were created for statistical analysis.

Table I: Acoustic Properties of Constituent Materials[33] and Water

| Material | Speed of Sound (m/s) | Attenuation (Np/m) | Density (kg/m$^3$) |
| --- | --- | --- | --- |
| Water | 1500 | 0 | 1000 |
| Red Marrow | 1450 | 12.55 | 1029 |
| Cortical Bone | 3514 | 54.55 | 1908 |

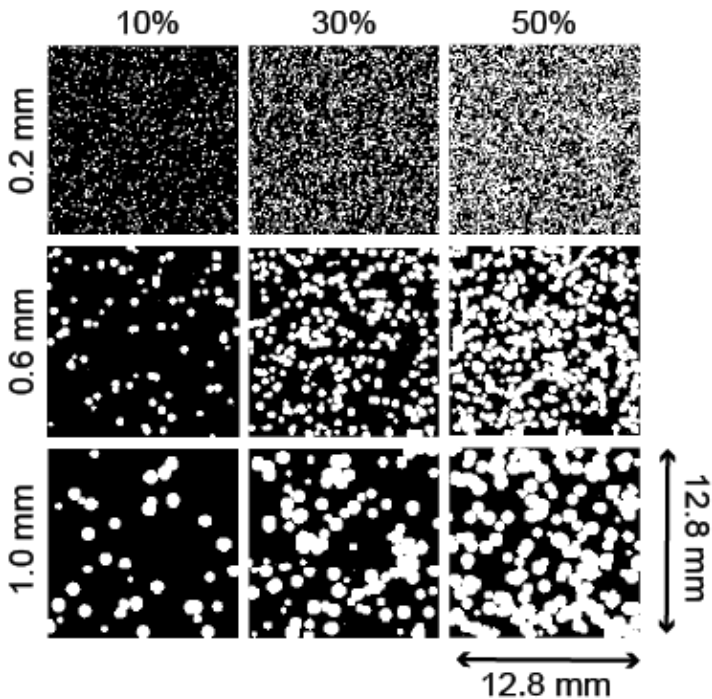

Figure 1: Phantom central transverse cross sections for pore diameters of 0.2 mm (top), 0.4 mm (middle), and 0.6 mm (bottom) at nominal porosities 10% (left), 30% (center), and 50% (right).

Additional phantoms with non-planar interfaces were constructed based on a clinical CT image of a human skull. The image was retrospectively obtained from an 86-year-old male patient treated for essential tremor. Three skull pieces corresponding to the temporal, frontal, and parietal bones were segmented from the skull CT. We then eroded 1 mm from the outer and inner tables to roughly remove the cortical layers while maintaining the shape. Finally, we added idealized skull microstructure by masking digital phantom volumes consisting of three porosities (10%, 30%, 50%) and three pore diameters (0.2 mm, 0.4 mm, 0.6 mm) for a total of 27 CT-based phantoms with non-planar skull/water interfaces. Further details of the CT-based phantom generation, including image acquisition and segmentation, are provided in the supplemental materials.

3D simulations
All simulations were completed using k-Wave version 1.4, a well-validated full-wave pseudospectral time-domain solver.[34,35] The extent of the grid was 12.8 mm x 12.8 mm x 7.6 mm with an isotropic spacing of 0.05 mm. Continuous uniform pressure sources (frequency = 650 kHz or 230 kHz, amplitude = 1.0 MPa, initial ramp = 4 cycles, phase = 0 rads) were propagated in the z-direction from the front plane of the grid. A 20-voxel-thick perfectly matched layer with 2.0 Np/voxel attenuation was added externally to the end plane to achieve an effective infinite domain without extending the computation grid. This addresses the assumed longitudinal periodicity in calculating the spatial derivatives with the Fourier Transform, enabling accurate and computationally efficient simulations.[36] Transverse matching layers were unnecessary as the assumed transverse periodicity causes the uniform source to remain uniform as it propagates (i.e., a plane wave as described in the k-Wave MATLAB examples[37]). The 5 mm thick phantoms were placed ten voxels into the grid, resulting in a water segment 0.5 mm in front and 2.1 mm in back of the phantom, as shown in Figure 2. Adding water increased the models' similarity to in vivo skull transmission, although the resulting interfaces between the water and phantoms, being planar, are non-physical. The pressure was measured in the last transverse plane (i.e., in water). The source pressure was smoothed by default within k-Wave using a Blackman filter to reduce high spatial frequency components. The Blackman filter is used by default because it offers a good compromise between reducing oscillations (low side lobes) with relatively minimal smoothing (narrow main lobe).[38] This filtering is included to avoid numerical errors due to discrete initial conditions causing non-physical oscillations.[39] The uniform source here is not strongly affected by this filtering. The medium properties, including speed of sound, attenuation, and density, were not smoothed. In our case, sharp boundaries were desired to simulate scattering accurately.[40] All simulations were run on a NVIDIA Tesla P40 GPU. For each phantom of a single porosity, pore diameter and randomization, a 650 kHz non-absorbing simulation took 3.6 minutes while an absorbing simulation took 23 minutes. A 230 kHz non-absorbing simulation took 4.4 minutes while an absorbing simulation took 86 minutes.

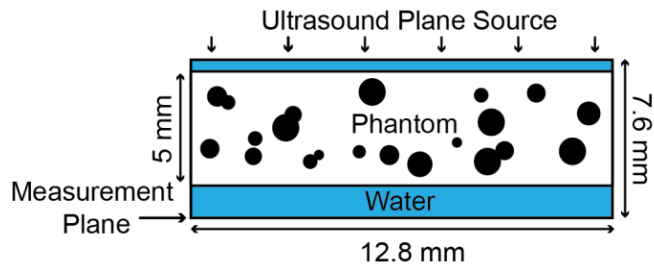

Figure 2: Longitudinal slice of a typical phantom, with the transverse source plane at the top propagating a steady-state wave through the phantom to the transverse measurement plane at the bottom.

The k-Wave sensor was defined to record the steady-state maximum pressure at all points in the measurement grid over the last two cycles. The simulation duration was set to 2.5 times the time of flight for 230 kHz and 2 times the time of flight for 650 kHz across the grid diagonal using the slowest speed of sound in the model (1450 m/s), giving a duration of 34 µs and 27 µs. The longer duration for lower frequencies accounts for the longer wavelength, which takes longer to reach a steady state. The time step was defined using a specified Courant-Friedrichs-Lewy (CFL) condition, where a smaller CFL condition results in a smaller time step.[36] The non-absorbing phantoms at both frequencies used a CFL of 0.3, while absorbing simulations used a CFL of 0.06 for 650 kHz sources and a CFL of 0.02 for 230 kHz sources.

Convergence testing was completed over the time duration and time step using the 9 phantoms shown in Figure 1, with representative pore diameters (0.2 mm, 0.6mm, 1.0 mm) and porosities (10%, 30%, 50%). The percentage difference of the mean steady-state pressure in the measurement plane with the test parameters compared to the reference parameters was used as a metric for convergence. For each test, we report the maximum percentage difference across the 9 test phantoms here, and the complete results are included in Supplemental Table 1. First, we increased the simulation's duration by 2 times to 68 µs at 230 kHz and 54 µs at 650 kHz, resulting in a maximum difference of 9.045% and 3.711%, respectively. We inferred time duration convergence for absorbing phantoms because the additional absorption will cause the system to approach a steady state faster. Next, the CFL was reduced by a factor of 2 to 0.2 in the non-absorbing simulations and 0.01 and 0.03 in the absorbing simulation at 230 kHz and 650 kHz. The maximum percentage difference was 0.113% and 0.165% in the non-absorbing simulations at 230 kHz and 650 kHz. The maximum percentage difference was 0.031% and 0.003% in the absorbing simulations at 230 kHz and 650 kHz. We did not conduct convergence testing in the grid spacing, as our grid spacing results in 44 points per wavelength in the component with the minimum velocity at 650 kHz, which is well above the recommended minimum of 4 points per wavelength.[40]

The loss mechanisms included in this study are intra-voxel absorption, inter-voxel scattering, and phase interference due to planar reflections and multiple scattering. Attenuation is included in k-Wave as loss mechanisms within a voxel, including intra-voxel absorption and intra-voxel scattering. k-Wave models both intra-voxel loss mechanisms as absorption only. This is consistent with attenuation values reported in the literature, which include some combination of absorption and scattering loss. Within this study, a non-absorbing simulation means no intra-voxel loss is included. We avoid calling these simulations "non-attenuating" because they still include some loss mechanisms, including inter-voxel scattering, which may be included in an attenuation measurement depending on the voxel resolution. The specific loss mechanisms through heterogeneous media are difficult to separate. As such, we report an insertion loss that includes scattering, phase interference, and absorption (for some cases).

The insertion loss is determined by the mean pressure transmitted through the phantom averaged over the field of view in the measurement plane, compared to a water-only reference, as follows:

$$Insertion\ Loss = 100 \times \left(1 - \frac{\bar{P}_{phantom}}{\bar{P}_{water}}\right) \qquad (1)$$

The simulations were completed under two conditions: non-absorbing and absorbing. For the non-absorbing case, we did not define an attenuation term in k-Wave, which is equivalent to using k-Wave's "no absorption" flag (this also results in no dispersion). Excluding absorption better emphasizes the portion of the loss caused by phase interference and inter-voxel scattering. This flexibility allows us to test our hypothesis that insertion loss depends on porosity and pore diameter.

Absorbing simulations, which then included a non-zero attenuation term, were completed to test the extent to which the addition of absorption impacts the insertion loss. Acoustic absorption in k-Wave is modeled by a frequency power law, which relates two loss terms to the frequency-dependent absorption.[41] These terms are the power law pre-factor and the frequency exponent. The exponent is related to the dispersion (speed-of-sound dependence on frequency) as required by the Kramers-Kronig relationship.[41] We assumed no dispersion, such that the speed of sound is independent of frequency. This assumption is appropriate for continuous, single-frequency sources. In the non-absorbing simulations, if no loss terms are defined, the default in k-Wave is no dispersion. In absorbing simulations, the absorption power law exponent can be set to 2, which also results in no dispersion. The absorption pre-factor must then be scaled by the square root of the ratio of the modeled frequency (230 kHz or 650 kHz) to 1.0 MHz to adjust for this exponent.

To investigate the effect of microstructure on the propagation of the wave through the different bone phantom configurations, we conducted simulations on various sets of these phantoms, including finding pressure patterns inside the phantoms and determining the insertion loss as a function of porosity and pore diameter, as described next.

RESULTS

We found the 3D pressure distributions for three representative non-absorbing phantoms with nominal 30% porosity and three different pore diameters at 650 kHz. Images of the results are shown in Figure 3, which shows the steady-state pressure patterns in the transverse and central longitudinal planes; the transverse patterns are found at the measurement plane in water. The longitudinal images include some bone structures where pressures are higher than in the transverse patterns, since the bone's higher acoustic impedance requires higher pressures than in water for the same propagating intensity. The pressure patterns are observed to be only weakly correlated with the location of the pore edges outlined by the white contour lines, particularly evident for the smaller pores. In Figure 3, the loss is listed above the transverse pressure patterns for each phantom. The insertion loss increases with pore diameter at this porosity. For example, the mean pressure in the transverse plane in A is 0.8 MPa; in C, it is 0.45 MPa, 1.8 times smaller.

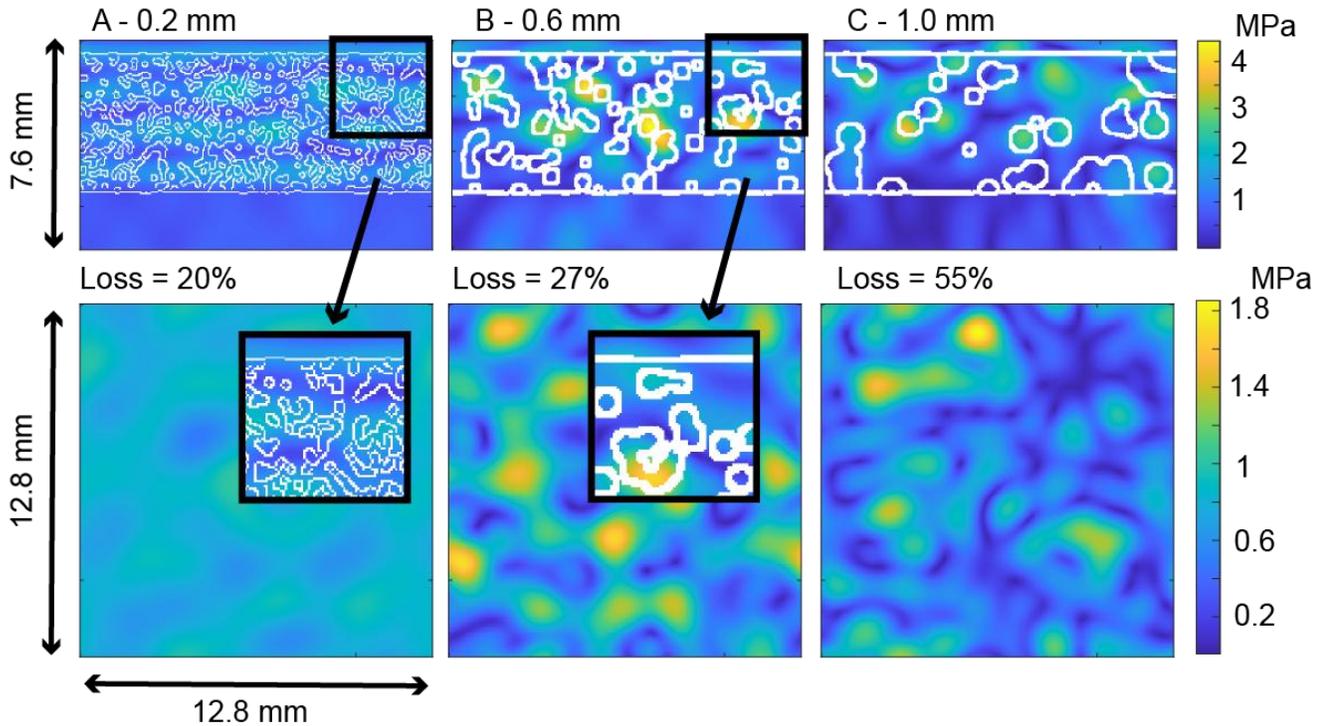

*Figure 3: Steady-state peak positive pressure patterns for three 30% porosity non-absorbing phantoms in central longitudinal (top row) and transverse (bottom row) views at 650 kHz. The transverse pressure distributions are shown at the measurement plane. The insets in A and B are included to show the fine microstructure, which is depicted by the white contour lines. The columns have pore diameters of: A) 0.2 mm, B) 0.6 mm, and C) 1.0 mm.*

The insertion loss as a function of porosity and pore size through a variety of non-absorbing phantoms at both frequencies is shown in Figure 4. The six lines represent phantoms with a single pore diameter and varying porosity. The error bars in the insertion loss represent the standard deviation of the loss across the five sets with different random pore distributions at each porosity and pore diameter. At 230 kHz, the loss depends on porosity but has minimal dependence on the microstructure. The greatest difference in loss between pore diameters is 6% at a nominal 20% porosity. The average percentage difference between the smallest and largest pore diameters is 1%.

At 650 kHz, the simulated insertion loss depends on both the porosity and the phantom's microstructure as characterized by the pore diameter. The loss for a given porosity tends to increase with the pore diameter (though a few exceptions to this trend exist in the low porosity region). Two relatively large differences in insertion loss as a function of pore diameter occur at nominal porosity values of 2.5% and 20%, corresponding to loss differences of 49% and 50%. The loss simulated at 0% porosity (no pores) corresponds to that of a purely cortical phantom, 28%. The loss above 50% porosity tends to converge to the value for a homogeneous marrow phantom, 0.01%. The two smaller pore diameter plots have two minima and one peak in the lower porosity region (<30%). As pore size increases, the loss difference between the peaks and minima fades until only a single peak occurs in the larger pore diameter plots.

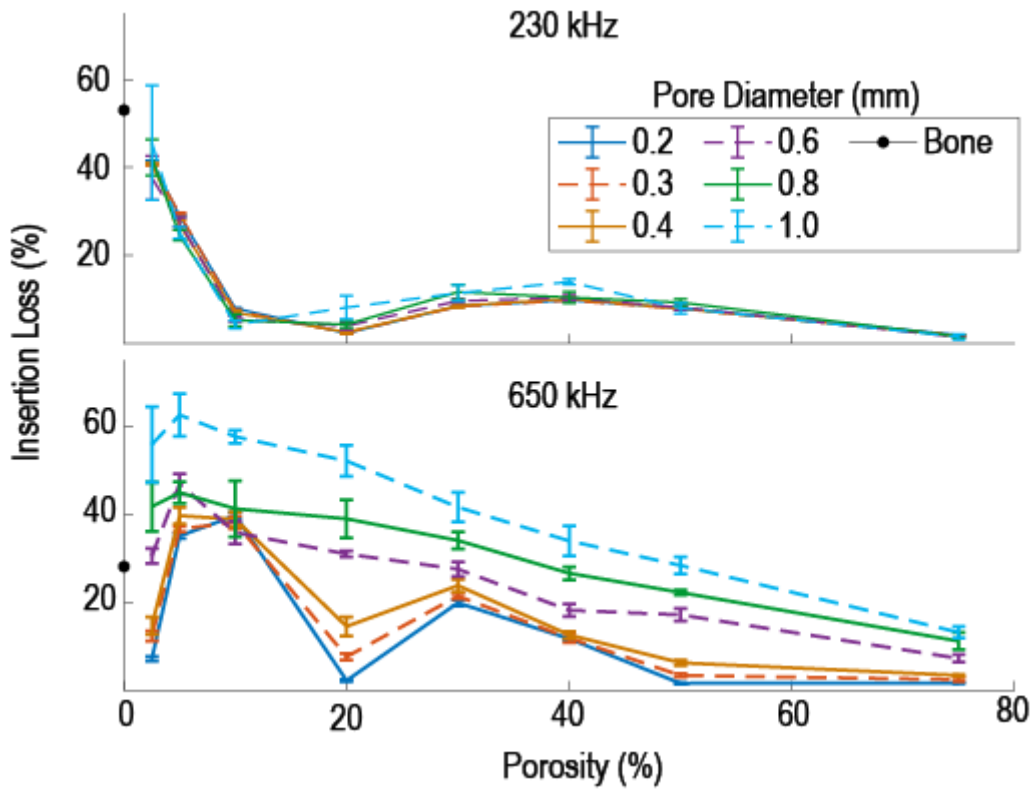

Figure 4: Insertion loss through non-absorbing phantoms as a function of porosity. Each line corresponds to a given pore diameter. Vertical error bars represent the standard deviation of the insertion loss across five sets of phantoms with different random pore placements. The data points are located on the mean porosity across the five sets.

Absorbing simulations were done to find the insertion loss through such phantoms with varying porosity and pore diameter. The results shown in Figure 5 have similar trends to those in the non-absorbing phantoms, but with increased loss for all cases. When averaged over all porosities and pore diameters, the loss increases by 2.9% at 230 kHz and 8.2% at 650 kHz when absorption is included. At 230 kHz, the largest increase occurs in the 0.4 mm pore diameter phantom at nominal 10% porosity, from 6.9% in the non-absorbing phantom to 13% in the absorbing phantom. At 650 kHz, the largest loss increase occurs in the 0.2 mm pore diameter phantom at 2.5% porosity, from 7.1% to 20.7%. Again, the main observation holds that the insertion loss depends on the porosity and the microstructure at 650 kHz. When compared to the non-absorbing phantom results (Figure 4), the largest noticeable change in the curves is seen for 0.2 mm pore diameters. In the non-absorbing phantoms, there is a variation of 37% in 0.2 mm pore diameter loss between the peak at 10% porosity and the minimum at 20% porosity; for the absorbing phantoms, this difference is reduced to 30%.

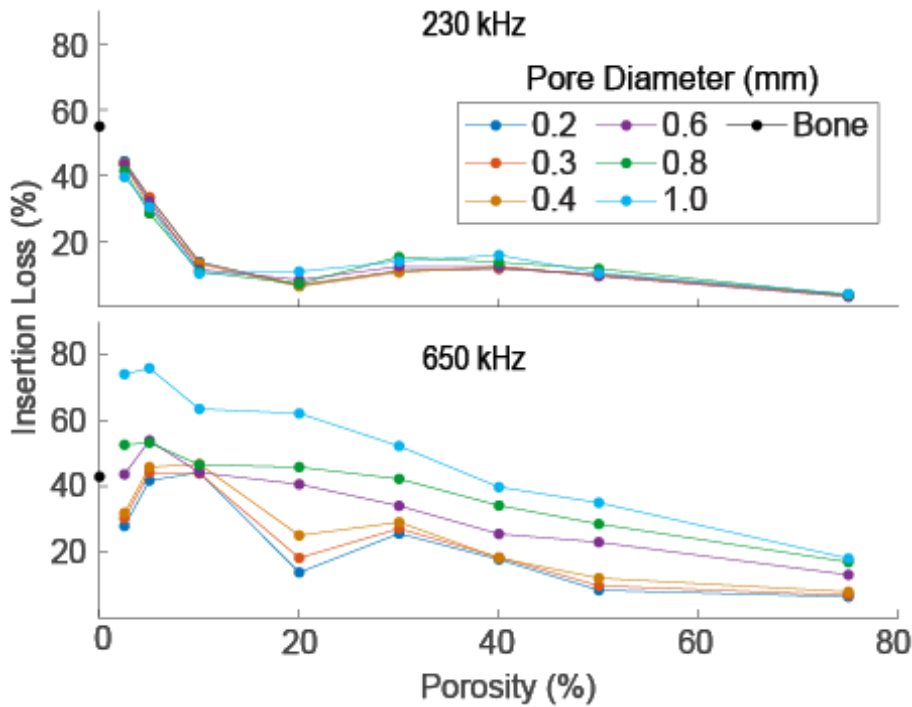

*Figure 5: Simulation of the insertion loss through phantoms of varying porosity and pore sizes including an absorbing component. Only a single randomization for each pore size and porosity was performed to minimize excessive computation time for absorbing phantoms. The loss increases an average of 8.2% compared to the non-absorbing results shown in Figure 4.*

The impact of multiple reflections on the insertion loss at 650 kHz was explored by plotting the mean pressure in each transverse plane along the propagation axis within four non-absorbing phantoms. We examined whether a standing wave pattern emerges and how that pattern depends on the microstructure. We used the four phantoms with two porosities and two pore diameters that showed contrasting patterns in the insertion loss in Figure 4. As such, we simulated nominal 20% porosity phantoms at both 0.2 mm and 1.0 mm pore diameters, where a large loss difference occurs between the two pore diameters, and nominal 10% porosity phantoms, where a smaller loss difference is seen. The results are shown in Figure 6. Evidence of standing waves is seen in the curves for the smaller diameter pore simulations, but not in the larger diameter curves. The pressures within a homogeneous bone phantom and water-only case are included for comparison.

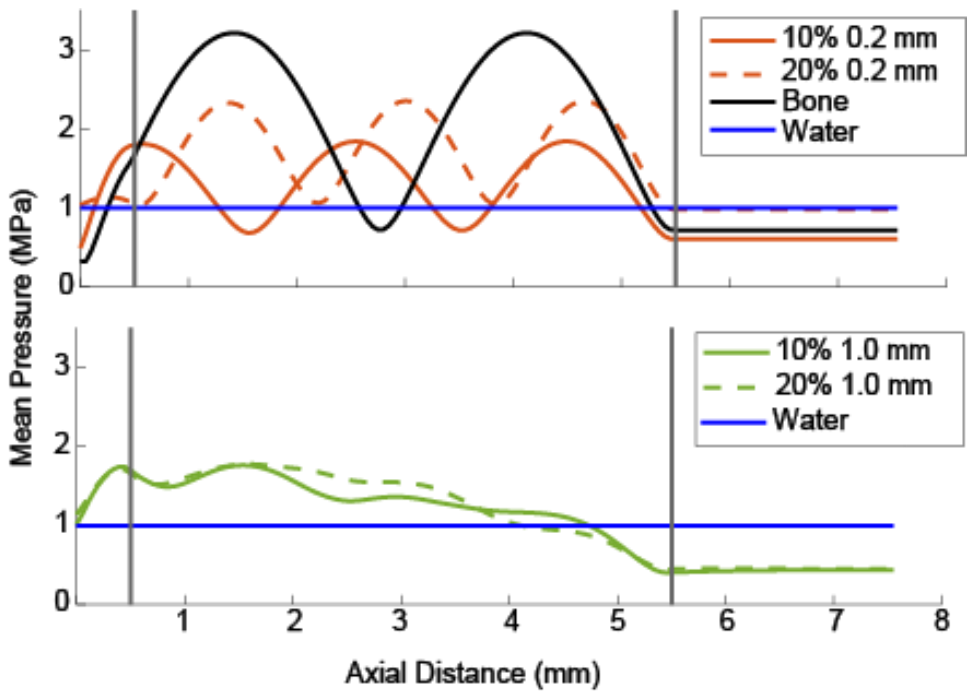

Figure 6: Mean pressure averaged over each transverse plane through 0.2 mm (top) and 1.2 mm (bottom) pore diameter phantoms at nominal 10% and 20% porosity with a 650 kHz source. Homogeneous bone and water-only curves are included for comparison. Vertical lines show the location of the two faces of the phantoms.

To more thoroughly investigate the effect that pore size has on the magnitude of the standing waves, we found the transverse pressure patterns through phantoms consisting of all six pore diameters with a nominal porosity of 20%, as shown in Figure 7. Clear standing waves are observed between the two planar faces within all three of the smaller pore diameter phantoms, but the amplitude of the standing waves decreases with increasing pore diameter. Note that the spacing between the peaks and valleys at the same nominal porosity slightly varies depending on the pore diameter. Within the two largest pore diameter phantoms, no prominent standing waves are observed.

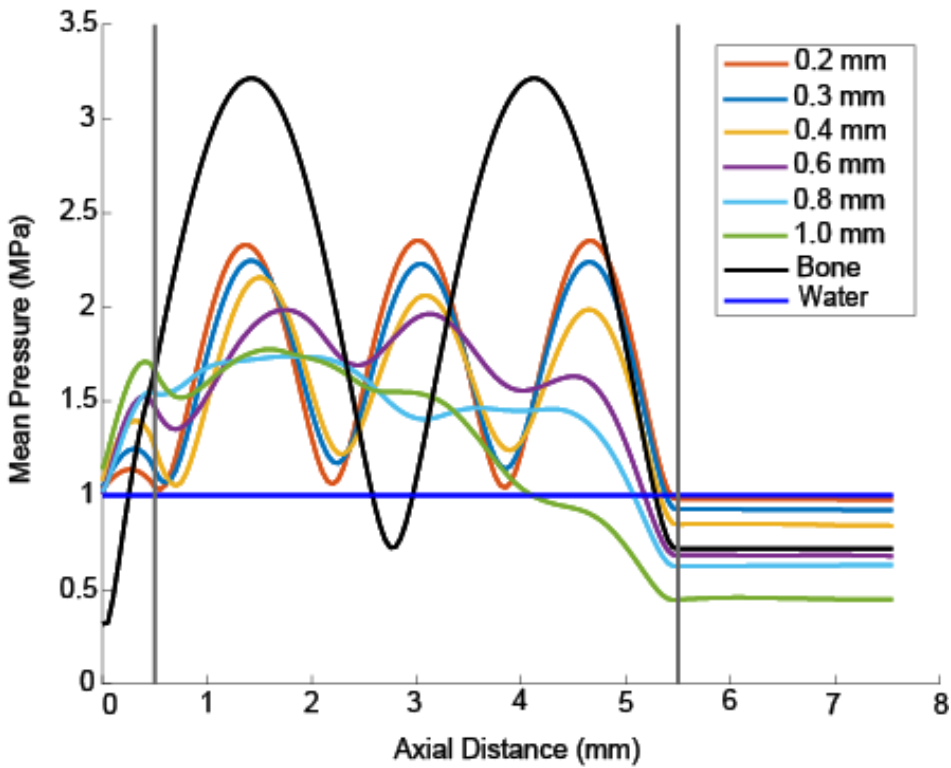

Figure 7: Mean pressure averaged over each transverse plane through 20% nominal porosity phantoms for six pore sizes with a 650 kHz source. The three smaller pore diameter phantoms exhibit clear standing waves. The 0.6 mm pore diameter line shows some standing wave effect. The two larger pore phantoms do not have standing wave effects. Vertical lines indicate the two faces of the phantoms. A water-only curve and a homogeneous bone curve are included for reference.

The microstructure's effect on digital phantoms that have non-planar water/bone interfaces based on real skull bone was examined. We used a 650 kHz planar source with the same simulation settings as above, except the grid size increased to include the larger phantoms. We found the 3D pressure distributions through each phantom type at 30% porosity and two pore sizes (0.2 mm and 0.6 mm). Images of the results are shown in Figure 8, which gives the steady-state pressure patterns in the transverse and central longitudinal planes; the transverse patterns are found in the measurement plane. The longitudinal images show the bone structure, including the curvature of the water/bone and bone/water interfaces and variable thickness within and between phantoms as indicated by the white contour lines. The transverse pressure patterns have an edge artifact seen as square patterns, which are more evident in the smaller pore size results. The edge artifact does not affect the insertion loss because we take an average over the measurement plane. The insertion loss, listed above the transverse pressure patterns for each phantom, increases with pore diameter at 30% porosity for each bone type. Similar to the phantoms with planar interfaces, we find that the insertion loss depends on pore diameter and increases as it increases.

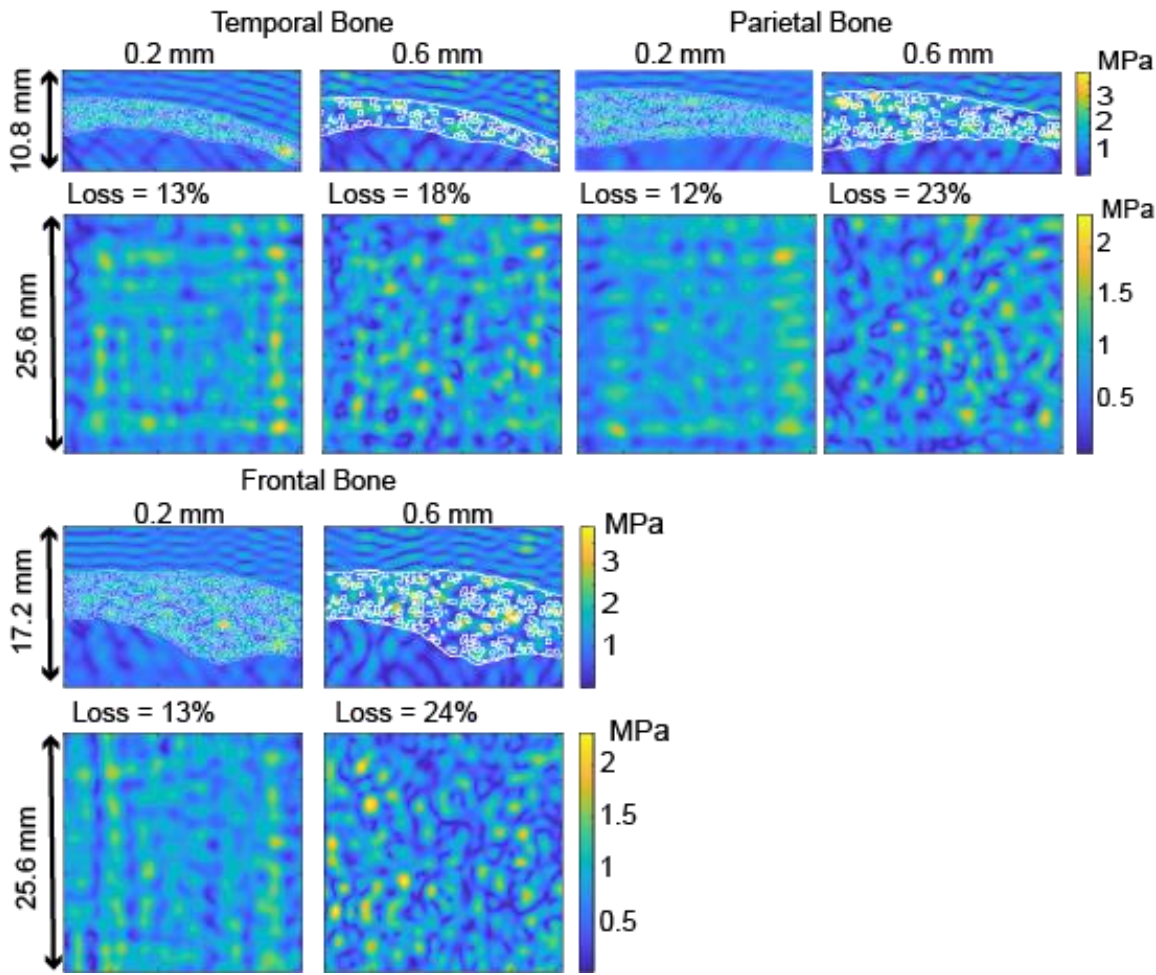

Figure 8: Steady-state peak positive pressure patterns through temporal, frontal, and parietal-shaped 30% porosity bone phantoms consisting of 0.2 mm (left) or 0.6 mm (right) pore diameters. The central longitudinal pressure patterns (top) show the phantom structure in white contours. Transverse pressure distributions at the measurement plane indicate that the insertion loss increases with pore diameter.

Figure 9 summarizes this result across the three bone types. As before, the insertion loss generally increases with pore size at each porosity. On average, the 0.6 mm pore diameter curve has the greatest loss (4.5%, 12.7%, and 7.5% in the temporal, frontal, and parietal bone types). The largest difference in loss between pore sizes at each porosity occurs in the frontal bone, which is thicker than the other bone types. An exception to this pore size trend occurs at 10% porosity in the parietal bone, with the 0.2 mm phantom having a 1.3% higher loss. The loss decreases with increasing porosity in all cases.

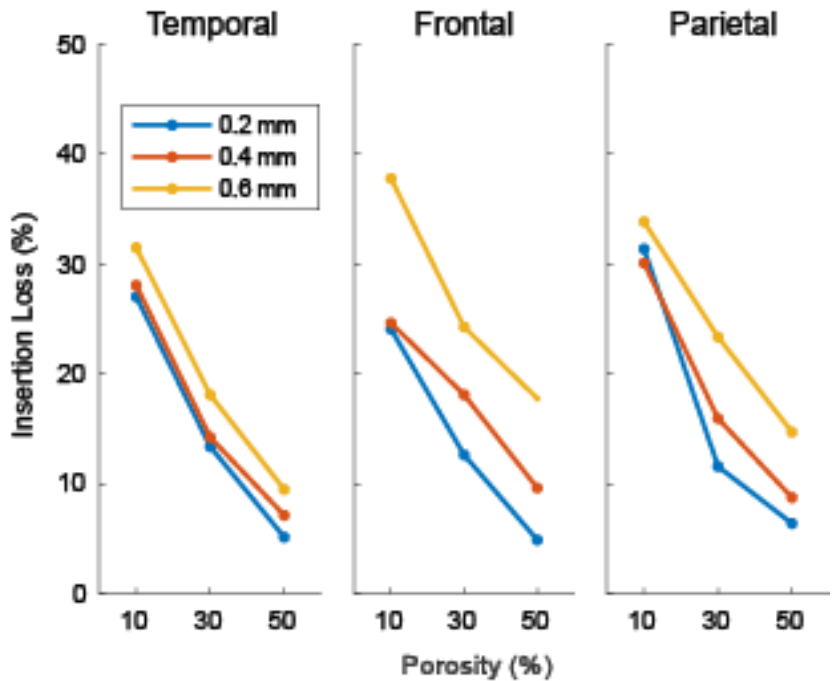

Figure 9: Insertion loss through temporal, frontal, and parietal bone-shaped phantoms consisting of three porosities and three pore diameters. Each line corresponds to a given pore diameter.

DISCUSSION
This study provides evidence that, at 650 kHz, the attenuation of skull bone depends on microstructure that is not resolved by clinical CT. We simulated the effect of microstructure on the loss of a wave's pressure as it propagates through a variety of digital bone phantoms with varying porosities and pore sizes. We found that the insertion loss, measured on a distal measurement plane, varied as a function of both porosity and pore diameter.

The results in Figure 3 can be interpreted as showing that the pressure distribution is strongly influenced by scattering from the phantom's surrounding microstructure. The spatial frequencies observed in the pressure patterns in the measurement plane correlate with the size of the pores within the microstructure. From Figure 3, the pattern within C shows higher spatial frequencies than A and B. This is likely due to increased scattering of waves into higher angles (higher spatial frequencies) by the coarser microstructures.

The primary result given in Figure 4 shows that the insertion loss depends on both the porosity and microstructure, as characterized by the pore diameter at 650 kHz. This conclusion also holds for absorbing phantoms, as shown in Figure 5. The general trends in Figures 4 and 5 follow scattering theory in the sense that larger pores result in more scattering and, thus, a higher insertion loss.[31,42] Minor exceptions to that trend can be explained by the complex interactions of overlapping and adjacent pores.[27,43] These interactions are difficult to quantify with the simplified assumptions of various scattering theories, but they are included in k-wave's full-wave approach.

Figures 4 and 5 also include loss results for a 230 kHz source, which depends on the porosity but is largely independent of the pore-size microstructure. This difference in loss behavior as a function of frequency may be attributed to a transition into the Rayleigh scattering regime at 230 kHz, when pore diameters less than 1 mm are smaller than the angular wavelength in bone ($\lambda_{bone}/2\pi$ = 2.43 mm/rad).[31,44] This result provides evidence that lower frequency sources are relatively insensitive to the microstructure, suggesting that a single relationship between attenuation and CT HUs may exist. Further, it provides additional consideration in the frequency tradeoff for a therapeutic transducer, as the microstructure may be less important when using lower frequency sources. This is consistent with a previous finding that the wave distortions are reduced at frequencies below 500 kHz; however, the lower frequency also increases the focal spot size and influence of standing waves.[45] Therefore, an improved understanding of the ultrasound interactions with microstructure will be most relevant for higher-frequency ultrasound applications.

It should be noted that plots relating bone attenuation to density and HUs reported in the acoustic literature are typically flipped horizontally compared to Figure 4, placing marrow (high porosity, low density) on the left and cortical bone (low porosity, high density) on the right with increasing HU. Accordingly, we have placed additional figures of our results that relate loss to density and HUs (assuming linear relationships of porosity, density, and HUs) in Figures 1-2 of the Supplemental Materials.[2,46] These figures scale the horizontal axis to relate the loss to density and HUs. The general trends observed with porosity and pore size are still evident.

Figures 6 and 7 indicate that standing waves between the two planar faces of the small pore diameter phantoms can affect the loss. Our planar phantoms are a simplified geometric approximation of bone with uniform thickness, improving the interpretability of the loss. However, the planar interfaces increase the effect of multiple coherent reflections, which also influences the loss. These standing waves are more pronounced for the relatively more homogeneous (small pore diameter) phantoms, possibly because increased scattering within the coarser phantoms disrupts the phase coherence of the reflected beams. Standing waves may explain the oscillations in the loss curves for the small pore sizes within the lower porosity regions in Figures 4 and 5, which are not seen for the larger pore sizes.

This standing wave effect is weaker in the non-planar interfaces of real skull bones, but—as shown in Figures 8 and 9—the effect of pore size on insertion loss persists in these more realistic bone shapes. The same trend with pore size is observed, with the loss generally higher through the larger pore diameter phantoms at the same porosity. The difference between the loss as a function of pore size changes between the three bones. Within the thicker bone (frontal), the wave passes through more of the microstructure, and as such, the relative loss between pore sizes increases due to the differences in scattering properties.

The results of this study are consistent with our hypothesis that the ultrasound loss through skull-mimicking phantoms depends not only on the porosity (and corresponding HU) but also on the microstructure at 650 kHz. This suggests that there is not a one-to-one correspondence between HU and acoustic loss. This lack of one-to-one correspondence may explain why prior studies have failed to find a consistent relationship between HU and attenuation.[6] Clinically, this means two patients with similar CT HU distributions could require different amplitude compensations to deliver the same acoustic intensity.

To overcome this uncertainty, we need to better characterize the ultrasound interactions with the microstructure. One approach is to improve the resolution of CT imaging. Super-resolution imaging with clinical CT and deep learning has been explored to enhance the imaging of bone microstructure for the diagnosis of osteoporosis.[47] Alternatively, photon-counting CT may offer improved imaging resolution, although this is not yet widely available.[48,49] Clinical CT has a nominal isotropic resolution of 0.5 mm, while photon-counting CT can achieve a nominal isotropic resolution of 0.2 mm.[50] Our results suggest this improvement in resolution may justify the increased expense of photon-counting CT. Another approach is to obtain micro-CTs of ex vivo skull bone flaps and perform through transmission measurements and simulations. Using an optimization routine, these measurements can be compared to determine the attenuation at a high resolution, which includes the microstructure effects.[3] However, the challenge remains to estimate these high-resolution acoustic properties from the clinically available CT.[51]

Beyond CT, a better understanding of the ultrasound interactions with skull bone microstructure may be facilitated with a few alternative approaches. Directly measuring the attenuation or loss may be possible using ultrasound backscatter or through-transmission measurements.[52,53] Both analytic and deep learning algorithms can extrapolate information about bone microstructure, including pore size and pore density, from the backscattered wave.[54–57] Alternatively, the scattering extinction length found from the backscattered wave can be related to the diffusion constant of a medium, and thereby the attenuation.[58] These measures may require a diagnostic ultrasound probe due to the hardware limitations of typical therapeutic transducers.

Knowledge of skull bone's porosities, pore shapes, and their associated scattering behavior may enable physics-based constraints.[57] Previous studies have used the relationships of scattering to porosity to constrain attenuation relationships to CT HUs.[10] One study proposed individualized attenuation relationships depending on the reported scattering attenuation relationship to porosity.[16] However, they conclude that a better understanding of scattering, especially within trabecular bone, could improve their model. Further, the sex, age, and skull location will likely influence the skull microstructure properties.[18,29,59] Knowledge of how these factors influence the loss may improve predictions.

Further work is needed to determine whether the relationship between phase velocity and HUs is also ill-determined. Two observations in this study suggest that the phase velocity for a given porosity does depend on pore diameter. First, the loss versus porosity curves in Figure 4 have minima at different locations for different pore diameters, suggesting their phase velocity depends on both microstructure and porosity. Second, the standing waves at a nominal 20% porosity, shown in Figure 7, have different spacing between peaks and valleys for different pore sizes. There is some evidence in the literature that phase velocity depends on the microstructure at frequencies relevant to therapeutic ultrasound.[60] At higher frequencies and using pulsed sources, it has been shown that the group and phase velocities depend on the microstructure and can be used to diagnose osteoporosis[61]

There are several limitations to our study. Our digital phantoms approximate real bones in order to demonstrate that the loss depends on small features that cannot be resolved with clinical CT, but each phantom employs a single-sized pore. Allowing the pores to overlap increases the phantom's similarity to the irregular pore shapes in real bone. However, a detailed comparison to a skull's true microstructure has not been completed. Thus, while our results show that microstructure does influence loss, they do not fully characterize the uncertainty in real bone.

We used a non-physical plane wave ultrasound source because it simplifies the interpretability of the scattering behavior in each phantom. Many ultrasound sources are quasi-planar spherical waves when they propagate through the skull, as the focus typically lies within the deeper brain regions. In cases where the field is more spatially irregular when entering the skull, either through near-field or focusing effects, the result is an increased dependence on the particular microstructure's local pore distribution. As such, we avoided near-field or focused sources because they would require a significant number of randomizations and simulations to characterize the global microstructure rather than the local microstructure. We used continuous ultrasound sources, which are appropriate for most therapeutic ultrasound applications that are quasi-continuous. Neuromodulation procedures are typically several milliseconds, while ablation procedures can be 10 to 20 seconds.[62,63] One type of histotripsy is an exception, as it uses sharp pulses; however, boiling histotripsy utilizes longer pulses in the 10-20 ms range.[64] The loss when using short pulses may be even more sensitive to the microstructure, as the dispersion that spreads the pulse envelope can result from microstructure interactions, as shown in prior quantitative ultrasound studies.[65,66]

Our simulation method makes some approximations, including using a fluid model based on an attenuation power law relationship that does not include viscoelastic interactions. This simplification is appropriate for normally incident sources.[67,68] Using a non-elastic simulation neglects mode conversion, which contributes negligible loss when high incident angles with the skull are avoided.[69] While viscoelastic interactions will likely affect the loss at bone/spherical pore interfaces, we expect these interactions to also depend on the relative size of the microstructure; as such, our conclusions are unlikely to change with the addition of viscoelastic interactions. We also assume a linear solution to the wave equation, which is appropriate for low pressure amplitudes. [35]

Conclusion

This study suggests that no single relationship exists at higher frequencies between ultrasound attenuation and clinical CT HUs since the insertion loss depends on the porosity and the skull microstructure, which typical clinical CT cannot resolve. This intrinsic uncertainty may contribute to the wide variability of reported attenuation relationships in literature. The skull microstructure strongly influences the scattering loss at 650 kHz, although the skull microstructure becomes less important at 230 kHz. Furthermore, coherent multiple reflections can further challenge the loss prediction. This uncertainty in CT-derived attenuation relationships leads to limitations in a simulation's ability to predict transcranial intensity. Accounting for this uncertainty in CT-derived acoustic simulations as a pretreatment planning tool should lead to safer and more effective future treatments.

Acknowledgments

Authors gratefully acknowledge funding from the Mark H. Huntsman Endowed Chair and NIH Grants F31DE032916, R21EB033117, and R01EB028316.

Conflict of Interest Statement

The authors have no relevant conflicts of interest to disclose.

Bibliography

1. Clement GT, Kullervo H. Correlation of Ultrasound Phase with Physical Skull Properties. *Ultrasound Med Biol*. 2002;28(5):617-624. doi:10.1016/S0301-5629(02)00503-3

2. Aubry JF, Tanter M, Pernot M, Thomas JL, Fink M. Experimental demonstration of noninvasive transskull adaptive focusing based on prior computed tomography scans. *J Acoust Soc Am*. 2003;113(1):84-93. doi:10.1121/1.1529663

3. Pinton G, Aubry J francois, Bossy E, Muller M, Pernot M. Attenuation , scattering , and absorption of ultrasound in the skull bone. *Med Phys*. Published online 2011. doi:10.1118/1.3668316

4. Pichardo S, Sin VW, Hynynen K. Multi-frequency characterization of the speed of sound and attenuation coefficient for longitudinal transmission of freshly excised human skulls. *Phys Med Biol*. 2011;56(1):219-250. doi:10.1088/0031-9155/56/1/014

5. Webb TD, Leung SA, Ghanouni P, Dahl JJ, Pelc NJ, Pauly KB. Acoustic Attenuation: Multifrequency Measurement and Relationship to CT and MR Imaging. *IEEE Trans Ultrason Ferroelectr Freq Control*. 2020;68(5):1532-1545. doi:10.1109/TUFFC.2020.3039743

6. Leung SA, Webb TD, Bitton RR, Ghanouni P, Butts Pauly K. A rapid beam simulation framework for transcranial focused ultrasound. *Sci Rep*. 2019;9(1):1-11. doi:10.1038/s41598-019-43775-6

7. Pichardo S. BabelBrain: An Open-Source Application for Prospective Modeling of Transcranial Focused Ultrasound for Neuromodulation Applications. *IEEE Trans Ultrason Ferroelectr Freq Control*. 2023;70(7):587-599. doi:10.1109/TUFFC.2023.3274046

8. Connor CW, Clement GT, Hynynen K. A unified model for the speed of sound in cranial bone based on genetic algorithm optimization. *Phys Med Biol*. 2002;47(22). doi:10.1088/0031-9155/47/22/302

9. Kyriakou A, Neufeld E, Werner B, Székely G, Kuster N. Full-wave acoustic and thermal modeling of transcranial ultrasound propagation and investigation of skull-induced aberration correction techniques: A feasibility study. *J Ther Ultrasound*. 2015;3(1). doi:10.1186/s40349-015-0032-9

10. Vyas U, Ghanouni P, Halpern CH, Elias J, Pauly KB. Predicting variation in subject thermal response during transcranial magnetic resonance guided focused ultrasound surgery: Comparison in seventeen subject datasets. *Med Phys*. Published online 2016.

11. McDannold N, White PJ, Cosgrove R. An element-wise approach for simulating transcranial MRI-guided focused ultrasound thermal ablation. *Phys Rev Res*. 2019;1(3). doi:10.1103/PhysRevResearch.1.033205

12. Webb TD, Leung SA, Rosenberg J, et al. Measurements of the Relationship between CT Hounsfield Units and Acoustic Velocity and How It Changes with Photon Energy and Reconstruction Method. *IEEE Trans Ultrason Ferroelectr Freq Control*. 2018;65(7):1111-1124. doi:10.1109/TUFFC.2018.2827899

13. Angla C, Larrat B, Gennisson JL, Chatillon S. Transcranial ultrasound simulations: A review. *Med Phys*. 2023;50(2):1051-1072. doi:10.1002/mp.15955

14. Attali D, Tiennot T, Schafer M, et al. Three-layer model with absorption for conservative estimation of the maximum acoustic transmission coefficient through the human skull for transcranial ultrasound stimulation. *Brain Stimul*. 2023;16(1):48-55. doi:10.1016/j.brs.2022.12.005

15. Robertson J, Martin E, Cox B, Treeby BE. Sensitivity of simulated transcranial ultrasound fields to acoustic medium property maps. *Phys Med Biol*. 2017;62(7):2559-2580. doi:10.1088/1361-6560/aa5e98

16. Tavakoli MB, Evans JA. The effect of bone structure on ultrasonic attenuation and velocity. *Ultrasonics*. 1992;30(6):389-395. doi:10.1016/0041-624X(92)90095-4

17. Mézière F, Muller M, Bossy E, Derode A. Measurements of ultrasound velocity and attenuation in numerical anisotropic porous media compared to Biot's and multiple scattering models. *Ultrasonics*. 2014;54(5):1146-1154. doi:10.1016/j.ultras.2013.09.013

18. Alexander SL, Rafaels K, Gunnarsson CA, Weerasooriya T. Structural analysis of the frontal and parietal bones of the human skull. *J Mech Behav Biomed Mater*. 2019;90(September 2018):689-701. doi:10.1016/j.jmbbm.2018.10.035

19. Martin RB, Ishida J. The relative effects of collagen fiber orientation, porosity, density, and mineralization on bone strength. *J Biomech*. 1989;22(5):419-426.

20. Rho JY, Hobatho MC, Ashman RB. Relations of mechanical properties to density and CT numbers in human bone. *Med Eng Phys*. 1995;17(5):347-355. doi:10.1016/1350-4533(95)97314-F

21. Zioupos P, Cook RB, Hutchinson JR. Some basic relationships between density values in cancellous and cortical bone. *J Biomech*. 2008;41(9):1961-1968. doi:10.1016/J.JBIOMECH.2008.03.025

22. O'Rahilly R, Muller F. *Basic Human Anatomy: A Regional Study of Human Structure*. Saunders; 1983.

23. Krieg MA, Barkmann R, Gonnelli S, et al. Quantitative Ultrasound in the Management of Osteoporosis: The 2007 ISCD Official Positions. *Journal of Clinical Densitometry*. 2008;11(1):163-187. doi:10.1016/j.jocd.2007.12.011

24. Bossy E, Padilla F, Peyrin F, Laugier P. Three-dimensional simulation of ultrasound propagation through trabecular bone structures measured by synchrontron microtomography. *Phys Med Biol*. 2005;50(23):5545-5556. doi:10.1088/0031-9155/50/23/009

25. Sasso M, Haïat G, Yamato Y, Naili S, Matsukawa M. Frequency Dependence of Ultrasonic Attenuation in Bovine Cortical Bone: An In Vitro Study. *Ultrasound Med Biol*. 2007;33(12):1933-1942. doi:10.1016/j.ultrasmedbio.2007.05.022

26. Brookes M, Revell WJ. Blood supply of flat bones. In: *Blood Supply of Bone*. Springer London; 1998:64-74. doi:10.1007/978-1-4471-1543-4_7

27. Yousefian O, White RD, Karbalaeisadegh Y, Banks HT, Muller M. The effect of pore size and density on ultrasonic attenuation in porous structures with mono-disperse random pore distribution: A two-dimensional in-silico study. *J Acoust Soc Am*. 2018;144(2):709-719. doi:10.1121/1.5049782

28. Wear KA. Ultrasonic scattering from cancellous bone: A review. *IEEE Trans Ultrason Ferroelectr Freq Control*. 2008;55(7):1432-1441. doi:10.1109/TUFFC.2008.818

29. Larsson E, Kajsa; U, G; T, Accardo A. *Morphological Characterization of the Human Calvarium in Relation to the Diploic and Cranial Thickness Utilizing X-Ray Computed Microtomography*. Vol 41. (Roa Romero LM, ed.). Springer International Publishing; 2014. doi:10.1007/978-3-319-00846-2

30. Faran JJ. Sound Scattering by Solid Cylinders and Spheres. *J Acoust Soc Am*. 1951;23(4):405-418. doi:10.1121/1.1906780

31. Sehgal CM. Quantitative relationship between tissue composition and scattering of ultrasound. *Journal of the Acoustical Society of America*. 1993;94(4):1944-1952. doi:10.1121/1.407517

32. Sehgal CM, Greenleaf JF. Scattering of ultrasound by tissues. *Ultrason Imaging*. 1984;6(1):60-80. doi:10.1177/016173468400600106

33. PA H, F DG, C B, et al. *IT'IS Database for Thermal and Electromagnetic Parameters for Biological Tissues*.; 2022. doi:10.13099/VIP21000-04-1

34. Treeby BE, Cox BT. k-Wave: MATLAB toolbox for the simulation and reconstruction of photoacoustic wave fields. *J Biomed Opt*. 2010;15(2):021314. doi:10.1117/1.3360308

35. Aubry J francois, Bates O, Boehm C, et al. Benchmark problems for transcranial ultrasound simulation : Intercomparison of compressional wave models. *The Journal of the Acoustic Society of America*. Published online February 10, 2022:1-18.

36. Tabei M, Mast TD, Waag RC. A k -space method for coupled first-order acoustic propagation equations . *J Acoust Soc Am*. 2002;111(1):53-63. doi:10.1121/1.1421344

37. Plane Wave Absorption Example. October 3, 2014. Accessed October 20, 2024. http://www.k-wave.org/documentation/example_ewp_plane_wave_absorption.php

38. Treeby B, Cox B, Jaros J. *K-Wave A MATLAB Toolbox for the Time Domain Simulation of Acoustic Wave Fields User Manual*.

39. Treeby BE, Cox BT. A k -space Green's function solution for acoustic initial value problems in homogeneous media with power law absorption . *J Acoust Soc Am*. 2011;129(6):3652-3660. doi:10.1121/1.3583537

40. Robertson JL, Cox BT, Treeby BE. Quantifying Numerical Errors in the Simulation of Transcranial Ultrasound using Pseudospectral Methods. *IEEE International Ultrasonics Symposium*. Published online 2014:2000-2003. doi:10.1109/ULTSYM.2014.0498

41. Treeby BE, Cox BT. Modeling power law absorption and dispersion for acoustic propagation using the fractional Laplacian. *J Acoust Soc Am*. 2010;127(5):2741-2748. doi:10.1121/1.3377056

42. Papadakis EP. Ultrasonic Attenuation Caused by Scattering in Polycrystalline Media. *Physical Acoustics*. 1968;4(PB):269-328. doi:10.1016/B978-0-12-395664-4.50018-3

43. Punurai W, Jarzynski J, Qu J, Kurtis KE, Jacobs LJ. Characterization of entrained air voids in cement paste with scattered ultrasound. *NDT&E International*. 2006;39(6):514-524. doi:10.1016/j.ndteint.2006.02.001

44. Omid Y, Yasamin K, Marie M. Frequency-dependent analysis of ultrasound apparent absorption coefficient in multiple scattering porous media: application to cortical bone. *Phys Med Biol*. 2021;66(3). doi:10.1088/1361-6560/abb934

45. MARQUET F, TUNG YS, KONOFAGOU EE. FEASIBILITY STUDY OF A CLINICAL BLOOD–BRAIN BARRIER OPENING ULTRASOUND SYSTEM. *Nano Life*. 2010;01(03n04):309-322. doi:10.1142/s1793984410000286

46. Marsac L, Chauvet D, La Greca R, et al. Ex vivo optimisation of a heterogeneous speed of sound model of the human skull for non-invasive transcranial focused ultrasound at 1 MHz. *International Journal of Hyperthermia*. 2017;33(6). doi:10.1080/02656736.2017.1295322

47. Rytky SJO, Tiulpin A, Finnilä MAJ, et al. Clinical Super-Resolution Computed Tomography of Bone Microstructure: Application in Musculoskeletal and Dental Imaging. *Ann Biomed Eng*. 2024;52(5):1255-1269. doi:10.1007/s10439-024-03450-y

48. Rajagopal JR, Schwartz FR, Solomon JB, Enterline DS, Samei E. High spatial-resolution skull base imaging with photon-counting CT and energy-integrating CT: a comparative phantom study. 2024;47(4):613-620. doi:10.1097/RCT.0000000000001464.High

49. Wehrse E, Sawall S, Klein L, Glemser P, Delorme S. Potential of ultra-high-resolution photon-counting CT of bone metastases : initial experiences in breast cancer patients. Published online 2019:1-8. doi:10.1038/s41523-020-00207-3

50. Benson JC, Rajendran K, Lane JI, et al. A New Frontier in Temporal Bone Imaging: Photon-Counting Detector CT Demonstrates Superior Visualization of Critical Anatomic Structures at Reduced Radiation Dose. *American Journal of Neuroradiology*. 2022;43(4):579-584. doi:10.3174/ajnr.A7452

51. Robertson J, Urban J, Stitzel J, Treeby BE. The effects of image homogenisation on simulated transcranial ultrasound propagation. *Phys Med Biol*. 2018;63(14):22-24. doi:10.1088/1361-6560/aacc33

52. Iori G, Du J, Hackenbeck J, Kilappa V, Raum K. Estimation of Cortical Bone Microstructure from Ultrasound Backscatter. *IEEE Trans Ultrason Ferroelectr Freq Control*. 2021;68(4):1081-1095. doi:10.1109/TUFFC.2020.3033050

53. Riis T, Feldman D, Losser A, Mickey B, Kubanek J. Device for Multifocal Delivery of Ultrasound Into Deep Brain Regions in Humans. *IEEE Trans Biomed Eng*. 2024;71(2):660-668. doi:10.1109/TBME.2023.3313987

54. Song Y, Kube CM, Turner JA, Li X. Statistics associated with the scattering of ultrasound from the microstructure. *Ultrasonics*. Published online 2017:2-5.

55. Karbalaeisadegh Y, Yousefian O, Muller M. Influence of Micro-Structural Parameters on Apparent Absorption Coefficient in Porous Structures Mimicking Cortical Bone. *IEEE International Ultrasonics Symposium*. 2018;2018-Octob:1-4. doi:10.1109/ULTSYM.2018.8579610

56. Mohanty K, Yousefian O, Karbalaeisadegh Y, Ulrich M, Grimal Q, Muller M. Artificial neural network to estimate micro-architectural properties of cortical bone using ultrasonic attenuation: A 2-D numerical study. *Comput Biol Med*. 2019;114(September):103457. doi:10.1016/j.compbiomed.2019.103457

57. White RD, Yousefian O, Banks HT, Alexanderian A, Muller M. Inferring pore radius and density from ultrasonic attenuation using physics-based modeling. *J Acoust Soc Am*. 2021;149(1):340-347. doi:10.1121/10.0003213

58. Brett Austin M, Kay R, Marie M. The Respective and Dependent Effects of Scatterring and Bone Matrix Absorption on Ultrasound Attenuation in Cortical Bone. *Phys Med Biol*. 2024;69(11). doi:10.1088/1361-6560/ad3fff

59. Chen H, Zhou X, Fujita H, Onozuka M, Kubo KY. Age-related changes in trabecular and cortical bone microstructure. *Int J Endocrinol*. 2013;2013. doi:10.1155/2013/213234

60. Webb TD, Fu F, Leung SA, et al. Improving Transcranial Acoustic Targeting: The Limits of CT-Based Velocity Estimates and the Role of MR. *IEEE Trans Ultrason Ferroelectr Freq Control*. 2022;69(9):2630-2637. doi:10.1109/TUFFC.2022.3192224

61. Grimal Q, Laugier P, eds. *Bone Quantitative Ultrasound*. Springer; 2022. doi:https://doi.org/10.1007/978-3-030-91979-5

62. Darmani G, Bergmann TO, Butts Pauly K, et al. Non-invasive transcranial ultrasound stimulation for neuromodulation. *Clinical Neurophysiology*. 2022;135:51-73. doi:10.1016/j.clinph.2021.12.010

63. Elias WJ, Huss D, Voss T, et al. A pilot study of focused ultrasound thalamotomy for essential tremor. *New England Journal of Medicine*. 2013;369(7):640-648. doi:10.1056/NEJMoa1300962

64. Sukovich JR, Cain CA, Pandey AS, et al. In vivo histotripsy brain treatment. *J Neurosurg*. 2019;131(4):1331-1338. doi:10.3171/2018.4.JNS172652

65. Droin P, eve Berger G, Laugier P. *Velocity Dispersion of Acoustic Waves in Cancellous Bone*. Vol 45.; 1998.

66. Wear KA. Group velocity, phase velocity, and dispersion in human calcaneus in vivo . *J Acoust Soc Am*. 2007;121(4):2431-2437. doi:10.1121/1.2697436

67. Treeby BE, Saratoon T. The contribution of shear wave absorption to ultrasound heating in bones: Coupled elastic and thermal modeling. In: *2015 IEEE International Ultrasonics Symposium, IUS 2015*. Institute of Electrical and Electronics Engineers Inc.; 2015. doi:10.1109/ULTSYM.2015.0296

68. Gao Y, Lauber B, Chen Y, Colacicco G, Razansky D, Estrada H. *Influence of Shear Waves on Transcranial Ultrasound Propagation in Cortical Brain Regions*.

69. White PJ, Clement GT, Hynynen K. Longitudinal and shear mode ultrasound propagation in human skull bone. *Ultrasonics*. Published online 2006. https://www.ncbi.nlm.nih.gov/pmc/articles/PMC3624763/pdf/nihms412728.pdf

Supplemental Materials

Digital Phantoms

This section describes how digital phantoms with varying porosity and pore size were generated. Our goal was to have a set of phantoms with a range of porosity from 0% to 75% comprised of different pore diameters. We chose six pore diameters (0.2 mm, 0.3 mm, 0.4 mm, 0.6 mm, 0.8 mm, and 1.0 mm), under a wavelength (650 kHz, 2.3 mm) in water. Because the pores overlap randomly, no one-to-one correspondence exists between the number of pores placed and the porosity. To overcome this, eight target porosities (2.5%, 5%, 10%, 20%, 30%, 40%, 50%, and 75%) were chosen and achieved through an iterative process. For each pore diameter, the number of pores corresponding to a 0.5% increase in porosity, assuming no overlap, was determined by supplemental equation 1.

$$N_{0.5\%} = 0.05 * \frac{3\,TV}{4\pi r^3} \qquad (S.1)$$

Then, this number of pores was randomly placed repeatedly, and the porosity was calculated after each iteration. The process was complete when the porosity was equal to or greater than the target porosity. The randomization was completed by a uniformly random vector permutation of vectors expressing every possible location within the grid. Then, pore locations were chosen from the permuted vector so that no location could be repeated.

The actual porosity could exceed the target porosity by a maximum of 0.5%, as that is the maximum number of pores placed at a time. Due to this slight variation, we will refer to the target porosities as nominal porosities. Five phantoms with different random pore positions were created for statistical analysis at each nominal porosity and pore size.

CT Phantoms

We constructed CT-based phantoms using segmented temporal, frontal, and parietal bone from a CT scan. A clinical CT skull image of an 86-year-old male treated for essential tremor was retrospectively obtained (University of Utah, IRB 00121352). The CT was acquired on a Siemens SOMATOM Edge Plus scanner with 120 kVp, an H60 kernel, 1 mm slice resolution, and 0.5 mm in-plane resolution.

The temporal, frontal, and parietal bone sections were manually segmented (approximately 40 mm × 40 mm × 20 mm) and resampled to 0.05 mm isotropic resolution, matching the previously used phantom resolution. The CTs were rotated to align the outer table parallel to the slice dimension. A threshold segmentation (>100 HUs) was applied to distinguish bone from air and brain, with any holes filled using MATLAB's imfill function, resulting in a solid bone-shaped mask. The lateral extents of the volume were cropped to 25.6 mm × 25.6 mm, ensuring the bone mask filled the volume. This corresponds to a lateral grid size of 512x512, which is computationally efficient within k-Wave. Finally, 1 mm of the outer and inner tables were removed using MATLAB's imerode function with a spherical kernel to reduce the cortical regions.

Microstructure was added with porosities of 10%, 30%, and 50%, consisting of pore diameters of 0.2 mm, 0.4 mm, and 0.6 mm. The previously defined phantoms (12.6 mm × 12.6 mm × 5 mm) were replicated to form a larger phantom volume, which was then masked using the segmented bone structure. The final phantom model retained curved interfaces based on real bone while having a defined microstructure.

Figure S1 shows central longitudinal slices through 30% phantoms consisting of 0.2 mm or 0.6 mm pore diameters.

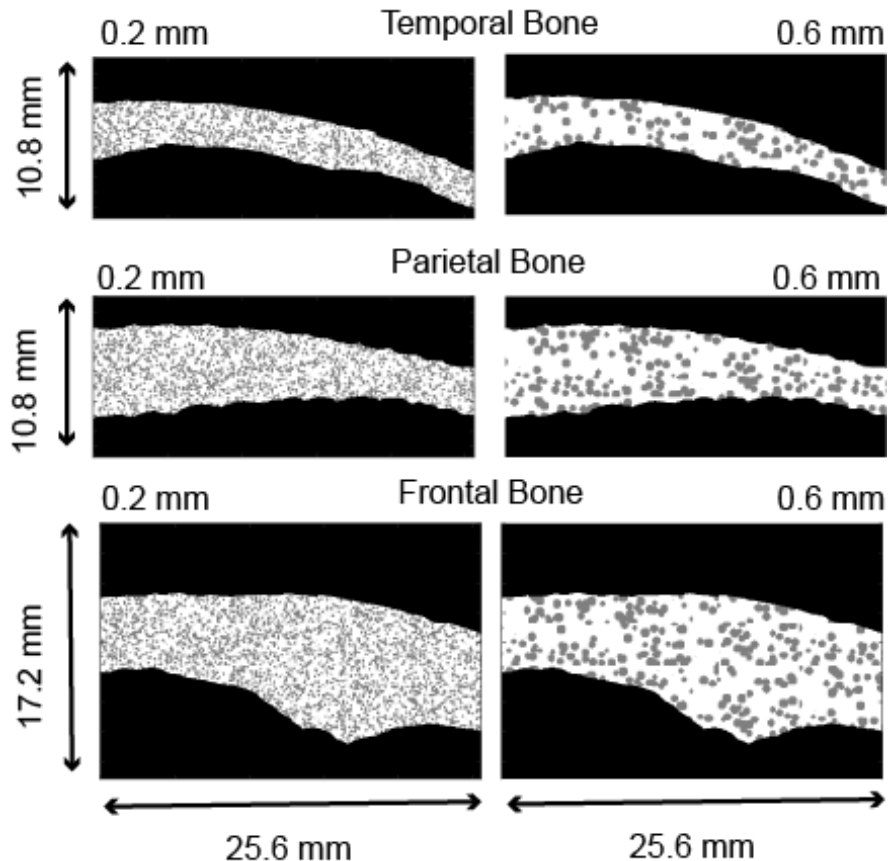

Supplemental Figure 1: CT-based phantoms with temporal, parietal, and frontal bone shapes. The idealized microstructure consists of spherical marrow pores in gray with 0.2 mm (left) or 0.6 mm (right) pore diameters

*randomly placed into a cortical background (white). The porosity is 30%. The black background corresponds to water.*

Convergence Testing

Convergence testing was completed over the time duration and time step using the 9 phantoms shown in Figure 1, with representative pore diameters (0.2 mm, 0.6mm, 1.0 mm) and porosities (10%, 30%, 50%). The percentage difference of the mean steady-state pressure in the measurement plane with the test parameters compared to the reference parameters was used as a metric for convergence. We report the maximum percentage difference across the nine test phantoms for each test in red. Our paper reports these maximum differences, which are all well below 10%, indicating good convergence in all cases. The full results are reported in Supplemental Table 1.

First, we increased the simulation's duration by 2 times to 68 µs at 230 kHz and 54 µs at 650 kHz, resulting in a maximum difference of 9.045% and 3.711%, respectively. We inferred time duration convergence for absorbing phantoms because the additional absorption will cause the system to approach a steady state faster. Next, the CFL was reduced by a factor of 2 to 0.2 in the non-absorbing simulations and 0.01 and 0.03 in the absorbing simulation at 230 kHz and 650 kHz. The maximum percentage difference was 0.113% and 0.165% in the non-absorbing simulations at 230 kHz and 650 kHz. The maximum percentage difference was 0.031% and 0.003% in the absorbing simulations at 230 kHz and 650 kHz. We did not conduct convergence testing in the grid spacing, as our grid spacing results in 44 points per wavelength in the component with the minimum velocity at 650 kHz, which is well above the recommended minimum of 4 points per wavelength.

Supplemental Table 1: Convergence Test Results Through Nine Representative Phantoms

| Phantom Porosity, Pore Diameter) | 230 kHz Non-Absorbing CFL | 650 kHz Non-Absorbing CFL | 230 kHz Absorbing CFL | 650 kHz Absorbing CFL | 230 kHz Non-Absorbing Duration | 650 kHz Non-Absorbing Duration |
|---|---|---|---|---|---|---|
| 10%, 0.2 mm | 0.055 | 0.006 | 0.017 | 0.002 | 2.467 | 0.386 |
| 30%, 0.2 mm | 0.028 | 0.031 | 0.031 | 0.001 | 2.697 | 0.731 |
| 50%, 0.2 mm | 0.027 | 0.164 | 0.031 | 0.003 | 0.733 | 3.711 |
| 10%, 0.6 mm | 0.086 | 0.022 | 0.018 | 0.003 | 3.308 | 1.739 |
| 30%, 0.6 mm | 0.024 | 0.005 | 0.017 | 0 | 3.665 | 1.865 |
| 50%, 0.6 mm | 0.03 | 0.014 | 0.026 | 0 | 9.045 | 1.763 |
| 10%, 1.0 mm | 0.113 | 0.024 | 0.017 | 0.001 | 0.065 | 2.941 |
| 30%, 1.0 mm | 0.02 | 0.009 | 0.012 | 0.001 | 2.622 | 0.932 |
| 50%, 1.0 mm | 0.026 | 0.022 | 0.023 | 0 | 8.113 | 0.623 |

Insertion Loss Vs. Density and Hounsfield Units

The primary result reported in Figure 4 shows the insertion loss through non-absorbing phantoms as a function of porosity. In literature, ultrasound loss (attenuation) is typically related to bone density or computed tomography (CT) Hounsfield Units (HU). Here, we report insertion loss as a function of density and HUs in supplemental Figures S2 and S3. This enables a more direct comparison to attenuation relationships found in the literature, as the linear transformations result in a horizontal axis flip, placing the low porosity phantoms at higher density and HU values.

In Figure S2, we assume density is linear with porosity, according to supplemental Equation S1 and as reported in Equation 4 in Aubry et al. 2003. We used the densities reported in Table 1 for marrow and cortical bone, 1029 kg/m$^3$ and 1908 kg/m$^3$, respectively. Other studies have used higher values for cortical bone density, for example, 2700 kg/m³ in Marsac et al. 2017; 2100 kg/m³ in Aubry et al. 2003; and 2200 kg/m³ in Marquet et al. 2009.

$$\rho = \emptyset \rho_{min} + (1 - \emptyset)\rho_{max} \tag{S1}$$

In Figure S3, we assume that the CT HUs is also linear with porosity and density, as described by supplemental Equation S2 and reported in Equation 4 in Marsac et al. 2017. In that study, they used a HU minimum of -1024 HU and a maximum of 2400 HU. The exact minimum and maximum HUs depend on the CT parameters and the individual skull density. Leung et al. 2019 reports several literature attenuation relationships in the range from 0 to 2000 HU. Here, we assume the same range as Leung et al. 2019.

$$\text{HU} = (HU_{max} - HU_{min})\frac{\rho - \rho_{min}}{\rho_{max} - \rho_{min}} + HU_{min} \tag{S2}$$

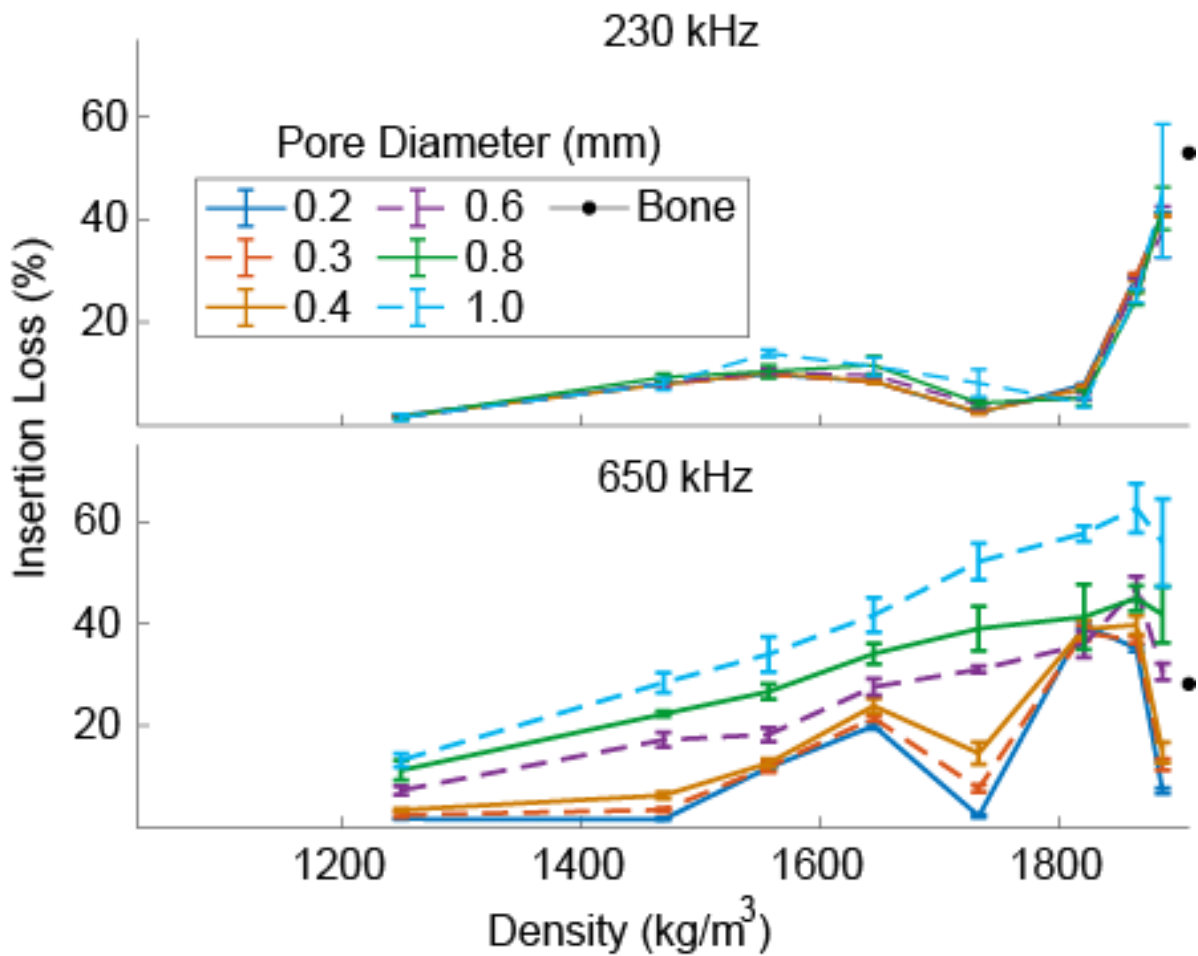

Supplemental Figure S2: Insertion loss through non-absorbing phantoms as a function of density assuming a linear relationship to porosity.

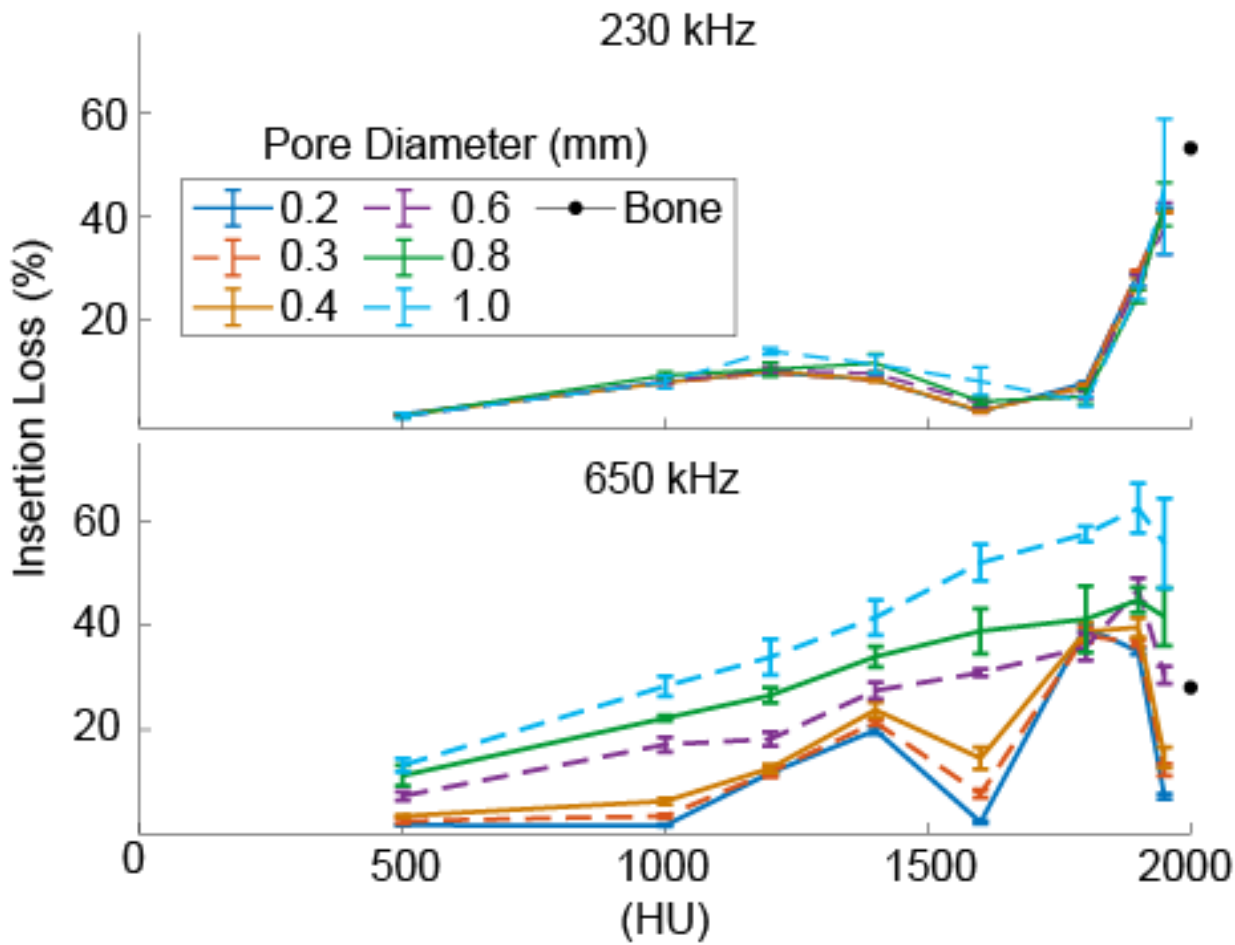

Supplemental Figure S3: Insertion loss through non-absorbing phantoms as a function of Hounsfield Units, assuming a linear relationship to density and porosity.

Supplemental References

Aubry, J.-F., Tanter, M., Pernot, M., Thomas, J.-L., & Fink, M. (2003). Experimental demonstration of noninvasive transskull adaptive focusing based on prior computed tomography scans. *The Journal of the Acoustical Society of America*, *113*(1), 84–93. https://doi.org/10.1121/1.1529663

Leung, S. A., Webb, T. D., Bitton, R. R., Ghanouni, P., & Butts Pauly, K. (2019). A rapid beam simulation framework for transcranial focused ultrasound. *Scientific Reports*, *9*(1), 1–11. https://doi.org/10.1038/s41598-019-43775-6

Marquet, F., Pernot, M., Aubry, J. F., Montaldo, G., Marsac, L., Tanter, M., & Fink, M. (2009). Non-invasive transcranial ultrasound therapy based on a 3D CT scan: Protocol validation and in vitro results. *Physics in Medicine and Biology*, *54*(9), 2597–2613. https://doi.org/10.1088/0031-9155/54/9/001

Marsac, L., Chauvet, D., La Greca, R., Boch, A. L., Chaumoitre, K., Tanter, M., & Aubry, J. F. (2017). Ex vivo optimisation of a heterogeneous speed of sound model of the human skull for non-invasive transcranial focused ultrasound at 1 MHz. *International Journal of Hyperthermia*, *33*(6). https://doi.org/10.1080/02656736.2017.1295322